\documentclass{article}
\usepackage{spconf,amsmath,graphicx,hyperref}

\usepackage{spconf}

\usepackage{amssymb,bm}
\usepackage{subfigure}
\usepackage{cite}
\usepackage{amsmath,amssymb}
\usepackage{multirow}
\usepackage{gensymb}
\usepackage{array}
\usepackage{enumerate}

\usepackage{colortbl}

\title{ 
Relative Transfer Matrix Estimator using
Covariance Subtraction
}
\name{%
Wageesha N. Manamperi$^{\dagger,\star}$, and 
Thushara D. Abhayapala$^\star$
}
\address{%
$^\dagger$ University of Moratuwa, Colombo, Sri Lanka \\
$^\star$ Australian National University, Canberra, Australia
}
\begin{document}
%
\maketitle
\begin{abstract}
The Relative Transfer Matrix (ReTM), recently introduced as a generalization of the relative transfer function for multiple receivers and sources, shows promising performance when applied to speech enhancement and speaker separation in noisy environments. 
Blindly estimating the ReTM of sound sources by exploiting the covariance matrices of multichannel recordings is highly beneficial for practical applications. 
In this paper, we use covariance subtraction to present a flexible and practically viable method for estimating the ReTM for a select set of independent sound sources.
To show the versatility of the method, we validated it through a speaker separation application under reverberant conditions.
Separation performance is evaluated at low signal-to-noise ratio levels $(\leq 0$ dB) in comparison with existing ReTM-based and relative transfer function–based estimators, in both simulated and real-life environments.
\end{abstract}

\vspace{-0.3cm}
\section{Introduction}
\label{sec:intro}
\vspace{-0.3cm}

Blindly estimating the relative transfer function (ReTF) without positional knowledge of receivers and sound source is highly attractive in practical applications such as robot audition, drone audition, teleconference, and hearing aids involving sound source localization, speech enhancement, speaker separation, acoustic echo cancellation \cite{nakadai2000active,wageeshadrtf,gannot2001signal,manamperi2024drone}.
Additionally, multiple simultaneously active sound sources are common in these applications and fail to hold W-disjoint orthogonality (WDO) \cite{yilmaz2004blind} for obtaining ReTFs of sound sources in mixtures. 
As a result, the generalization of the ReTF for multiple sound sources introduced as the relative transfer matrix (ReTM) \cite{abhayapala2023generalizing} is significantly beneficial in the field of audio signal processing. 
This paper presents a method to estimate the ReTM for a selected combination of active sources using the covariance subtraction property of independent signals.

Many approaches to estimate ReTF for a single active sound source have been developed 
\cite{shalvi1996system,cohen2004relative,middelberg2023relative,serizel2014low,varzandeh2017iterative,markovich2009multichannel}
Among them, covariance-based methods have gained attention due to their practical feasibility as well as simplicity \cite{middelberg2023relative,serizel2014low,varzandeh2017iterative,markovich2009multichannel}.
ReTF for a multiple and concurrent source scenario has been proposed \cite{schwartz2017two,koutrouvelis2019robust,li2022low,dietzen2020square,cherkassky2019successive}. Common drawbacks to these approaches are assuming WDO conditions, and a sufficiently high signal-to-background noise ratio (SNR).

Recently, two techniques have been proposed to generalize the ReTF for multiple simultaneous sound sources \cite{deleforge2015towards,abhayapala2023generalizing}.
In \cite{deleforge2015towards}, the ReTF generalization was proposed to localize multiple sources, where the number of active sources and (either full or partially) activation of all sources were assumed for estimation. 

In \cite{abhayapala2023generalizing}, the ReTM was introduced, which does not require source counting for speech enhancement or speaker separation in multi-source noisy reverberant environments, and subsequent works \cite{abhayapala2024,wnmanamperiReTM2ReTF,wnmanamperiReTMdDrone,wnmanamperiReTMSSS} have demonstrated its promising performance. 
Similar to ReTF, ReTM is independent of the signals emitted by multiple sources. By dividing multiple microphones into two groups, we can blindly estimate the ReTM from the received signals using the covariance-based method \cite{abhayapala2023generalizing}. 
However, in practice, estimating the ReTM from segments where all desired sources are inactive in low-SNR scenarios is challenging \cite{wnmanamperiReTMSSS}. Alternatively, this paper proposes a practically viable ReTM estimator for undesired sound sources using covariance subtraction \cite{serizel2014low,varzandeh2017iterative,middelberg2023relative}
which estimates the ReTM with respect to a select group of sources to handle overlapping and continuously active sources.
The proposed method's applicability for speaker separation in a typical use case of a meeting scenario, is evaluated by referring to our previous work \cite{wnmanamperiReTMSSS} and compared with the ReTF-based estimator using extensive experimental results in terms of speech intelligibility, signal-to-interference ratio, and signal-to-distortion ratio at very low SNR levels.

%
\vspace{-0.3cm}

\section{The Relative Transfer Matrix}\label{sec:2}

\vspace{-0.3cm}
\subsection{Signal model}\label{sec:2.1}
\vspace{-0.3cm}

Let us consider a reverberant environment with $\mathcal{L}$ sound sources (both speech $(S)$ and background noise $(N)$ sources). In the short time Fourier transform (STFT) domain, we denote $S_\ell(f,t) $, $\ell = {1,\cdots,\mathcal{L}}$ as the source signal.

Let there be $Q$ arbitrary distributed microphones in the room. We divide them to two groups of microphones, $\{A\}$ and $\{B\}$ with $Q_A$ and $Q_B$ microphones, respectively ($Q = Q_A + Q_B$). We denote $\mathbf{M}_A(f,t) $ and $\mathbf{M}_B(f,t) $  as the vector of received signals at microphone groups A and B, respectively. Then the received signals at each microphone group in matrix form as
%
\begin{equation}\label{eqn:M_A}
    \mathbf{M}_A(f,t) = \mathbf{H}_A(f)\mathbf{S}(f,t),  \mbox{ and}
\end{equation}
%
\begin{equation}\label{eqn:M_B}
    \mathbf{M}_B(f,t) = \mathbf{H}_B(f)\mathbf{S}(f,t), \mbox{ where}
\end{equation}
\[
\mathbf{S}(f,t) =  [S_1, \ldots, S_\ell, \ldots, S_{\mathcal{L}}]^T
= \begin{bmatrix}
\mathbf{S}^{(S)}(f,t) &
\mathbf{S}^{(N)}(f,t) 
\end{bmatrix}^T
\]
 and $\{\cdot\}^T$ is  the matrix transpose. Note that $\mathbf{S}^{(S)}$ and $\mathbf{S}^{(N)}$ are the vectors of speech and background noise source signals, respectively. Also, $\mathbf{H}_A(f) \in \mathbb{C}^{Q_A \times \mathcal{L}}$ and $\mathbf{H}_B(f) \in \mathbb{C}^{Q_B \times \mathcal{L}}$ are the matrices with elements defined by the acoustic transfer functions, and defined as $\mathbf{H}_A(f) = [\mathbf{H}^{(S)}_A(f) ~ \mathbf{H}^{(N)}_A(f)]$, and $\mathbf{H}_B(f) = [\mathbf{H}^{(S)}_B(f) ~ \mathbf{H}^{(N)}_B(f)]$.

The relative transfer matrix (ReTM) with respect to all active sources, $\bm{\mathcal{R}}_{AB}(f)$, is defined as in \cite{abhayapala2023generalizing} \begin{equation}\label{eqn:retm_hahb}
    \bm{\mathcal{R}}_{AB}(f) \triangleq \mathbf{H}_A(f) \mathbf{H}_B(f)^\dagger,
\end{equation}
where $(\cdot)^\dagger$ is Moore-Penrose inverse, assuming the validity, i.e., $Q_B \geq \mathcal{L}$. Thus, we can relate the received signal at group $\{A\}$ and $\{B\}$ using
%
\begin{equation} \nonumber 
     \mathbf{M}_A(f,t)  = \bm{\mathcal{R}}_{AB}(f) \, \mathbf{M}_B(f,t).
\end{equation}
%
Note that $\bm{\mathcal{R}}_{AB}(f)$ 
is defined by the spatial properties of the sound sources and the acoustic environment such that it is independent of the sound source signals. In applications, the ReTM is invariant for a wide-sense stationary (WSS) acoustic environment.

\vspace{-0.3cm}
\subsection{ReTM estimation with covariance matrices}\label{sec:2.2}
\vspace{-0.3cm}

In practice, the ReTM of the sound sources can be estimated using a segment of the microphone recording using the covariance matrices-based approach \cite{abhayapala2023generalizing}, given as 
\begin{equation}\label{eqn:retm_phi}
     \bm{\mathcal{R}}_{AB}(f) \approx \bm{\mathcal{P}}_{AA}(f) \,  \Big( \bm{\mathcal{P}}_{BA}(f) \Big)^\dagger, \mbox{ where}
\end{equation}
\begin{equation}\label{eqn:phirec_AA}
    \bm{\mathcal{P}}_{AA}(f) \triangleq E\{\mathbf{M}_A(f,t) \mathbf{M}_A^*(f,t)\}, 
\end{equation}
\begin{equation}\label{eqn:phirec_BA}
    \bm{\mathcal{P}}_{BA}(f)  \triangleq E\{\mathbf{M}_B(f,t) \mathbf{M}_A^*(f,t)\},
\end{equation}
$E\{\cdot\}$ denotes the expectation which can be obtained by averaging across the time frames, and $[\cdot]^*$ denotes the conjugate transpose. 
%
To make the exposition concise, we omit the dependency of time ($t$) and frequency ($f$) in the following sections. 
%
This paper aims to develop a practical ReTM estimator 
that does not require the segments of 
a set of 
sources to be inactive, with covariance subtraction, 
which we propose next.


\vspace{-0.3cm}
\section{Proposed Practical ReTM Estimator using Covariance Subtraction}\label{sec:3}
\vspace{-0.3cm}
This section uses covariance matrices to propose a method for estimating the ReTM of a selected set of individual sound sources using covariance subtraction.


Expanding \eqref{eqn:phirec_AA} by substituting \eqref{eqn:M_A} as
\begin{align}
\bm{\mathcal{P}}_{AA} 
 & = E\{ 
\scalebox{0.9}{$
\begin{bmatrix}
\mathbf{H}^{(S)}_A &  \mathbf{H}^{(N)}_A
\end{bmatrix}  
\begin{bmatrix}
\mathbf{S}^{(S)} \\
\mathbf{S}^{(N)} 
\end{bmatrix} 
\begin{bmatrix}
\mathbf{S}^{(S)^*} & \mathbf{S}^{(N)^*} 
\end{bmatrix} 
\begin{bmatrix}
\mathbf{H}^{(S)^*}_A \\
\mathbf{H}^{(N)^*}_A 
\end{bmatrix} 
$}
\}, \nonumber \\
 & =   
\begin{bmatrix}
\mathbf{H}^{(S)}_A &  \mathbf{H}^{(N)}_A
\end{bmatrix}
\begin{bmatrix}
\bm{\mathcal{P}}_{SS} &  \mathbf{0}_{Q_A \times Q_A} \\
\mathbf{0}_{Q_A \times Q_A} & \bm{\mathcal{P}}_{NN} 
\end{bmatrix}
\begin{bmatrix}
\mathbf{H}^{(S)^*}_A \\
\mathbf{H}^{(N)^*}_A 
\end{bmatrix}, \nonumber \\
 &  = 
\begin{bmatrix}
\mathbf{H}^{(S)}_A \bm{\mathcal{P}}_{SS} &  
\mathbf{H}^{(N)}_A \bm{\mathcal{P}}_{NN} 
\end{bmatrix}
\begin{bmatrix}
\mathbf{H}^{(S)^*}_A \\
\mathbf{H}^{(N)^*}_A 
\end{bmatrix}
, \nonumber \\
 &  = 
\underbrace{\mathbf{H}^{(S)}_A \bm{\mathcal{P}}_{SS} \mathbf{H}^{(S)^*}_A}_{\bm{\mathcal{P}}_{AA}^{(S)}} + 
\underbrace{\mathbf{H}^{(N)}_A \bm{\mathcal{P}}_{NN} \mathbf{H}^{(N)^*}_A.}_{\bm{\mathcal{P}}_{AA}^{(N)}}
\label{eqn:phi_AA}
\end{align}
where 
$
\bm{\mathcal{P}}_{SS} = E\{\mathbf{S}^{(S)} \mathbf{S}^{(S)^*}\}
$, and $
\bm{\mathcal{P}}_{NN} = E\{\mathbf{S}^{(N)}  \mathbf{S}^{(N)^*} \}
$ are the 
auto-covariance matrices of the speech and noise source signals, respectively.

Similarly, substituting \eqref{eqn:M_B} into \eqref{eqn:phirec_BA}, we can write
\begin{align}
\bm{\mathcal{P}}_{BA} 
 &  = 
\begin{bmatrix}
\mathbf{H}^{(S)}_B \bm{\mathcal{P}}_{SS} &  
\mathbf{H}^{(N)}_B \bm{\mathcal{P}}_{NN} 
\end{bmatrix}
\begin{bmatrix}
\mathbf{H}^{(S)^*}_A \\
\mathbf{H}^{(N)^*}_A 
\end{bmatrix}
, \nonumber \\
 &  = 
\underbrace{\mathbf{H}^{(S)}_B \bm{\mathcal{P}}_{SS} \mathbf{H}^{(S)^*}_A}_{\bm{\mathcal{P}}_{BA}^{(S)}} + 
\underbrace{\mathbf{H}^{(N)}_B \bm{\mathcal{P}}_{NN} \mathbf{H}^{(N)^*}_A.}_{\bm{\mathcal{P}}_{BA}^{(N)}}
\label{eqn:phi_BA}
\end{align}

Rearranging \eqref{eqn:phi_AA} gives an expression for the speech sources’ covariance matrix of microphone group $\{A\}$ in terms of the covariance matrices of the received signal and the noise source component of the received signal,
\begin{equation}\label{eqn:phiS_AA}
\bm{\mathcal{P}}_{AA}^{(S)} = \bm{\mathcal{P}}_{AA} - 
\bm{\mathcal{P}}_{AA}^{(N)}.
\end{equation}

From \eqref{eqn:phi_BA}, the speech sources’ covariance matrix of microphone groups $\{A\}$ and $\{B\}$ can be estimated as 
\begin{equation}\label{eqn:phiS_BA}
\bm{\mathcal{P}}_{BA}^{(S)} = \bm{\mathcal{P}}_{BA} - 
\bm{\mathcal{P}}_{BA}^{(N)}.
\end{equation}

Let $\bm{\mathcal{R}}_{AB}^{(S)}$ be the ReTM of the speech sources. Substituting \eqref{eqn:phiS_AA}, and \eqref{eqn:phiS_BA} back into \eqref{eqn:retm_phi} provides the ReTM of speech sources as
\begin{equation}\label{eqn:retm_s}
\bm{\mathcal{R}}_{AB}^{(S)} = 
\Big(\bm{\mathcal{P}}_{AA} - 
\bm{\mathcal{P}}_{AA}^{(N)} \Big)
\,  \Big( \bm{\mathcal{P}}_{BA} - 
\bm{\mathcal{P}}_{BA}^{(N)} \Big)^\dagger.
\end{equation}
We can observe from \eqref{eqn:retm_s} that the ReTM of the speech sources can be estimated by using the covariance matrices of the received signals and the received noise-only signals. This allows us to confirm the ReTM of speech sources can be estimated using covariance subtraction.

We note that the ReTM of the background noise sources $\bm{\mathcal{R}}_{AB}^{(N)}$ can be blindly estimated from the received noise-only signals (no active speech) as 
\begin{equation}\label{eqn:retm_n}
\bm{\mathcal{R}}_{AB}^{(N)} = 
\bm{\mathcal{P}}_{AA}^{(N)} \,  \Big(\bm{\mathcal{P}}_{BA}^{(N)}\Big)^\dagger.
\end{equation}
Furthermore, rearranging \eqref{eqn:retm_phi} (by substituting both \eqref{eqn:phi_AA} and \eqref{eqn:phi_BA}) gives an expression for the ReTM of all sound sources in
terms of the covariance matrices of the speech and noise sources between microphone groups $\{A\}$ and $\{B\}$ as
\begin{equation}\label{eqn:retm_total}
\bm{\mathcal{R}}_{AB} = 
\Big(\bm{\mathcal{P}}_{AA}^{(S)} + 
\bm{\mathcal{P}}_{AA}^{(N)} \Big)
\,  \Big( \bm{\mathcal{P}}_{BA}^{(S)} +  
\bm{\mathcal{P}}_{BA}^{(N)} \Big)^\dagger.
\end{equation}
We can observe from \eqref{eqn:retm_total} that, unlike the covariance matrices of the speech and noise sources, the ReTM of the speech and noise sources cannot be added together. Hence, ReTMs are not additive. However, we observe that covariance matrices can be manipulated to derive the ReTM of the speech sources. This property then allows for the generalization of ReTM estimation to be calculated by a select set of independent sound sources.

Following the simplifications in \eqref{eqn:phi_AA} and \eqref{eqn:phi_BA}, we can decompose the covariance matrices of independent $\mathcal{L}$ sources between microphone groups $\{A\}$ and $\{B\}$ into its individual source covariance matrices,  $\bm{\mathcal{P}}_{AA}^{(\ell)}$ and $\bm{\mathcal{P}}_{BA}^{(\ell)}$, 
 expressed as 
\begin{equation}\label{eqn:phiAA_generalization}
    \bm{\mathcal{P}}_{AA}  = \sum_{\ell=1}^\mathcal{L}  \mathbf{h}^{(\ell)}_A \mathcal{P}_{\ell} \mathbf{h}^{(\ell)^*}_A = \sum_{\ell=1}^\mathcal{L} \bm{\mathcal{P}}_{AA}^{(\ell)},
\end{equation}
\begin{equation}\label{eqn:phiBA_generalization}
    \bm{\mathcal{P}}_{BA}  = \sum_{\ell=1}^\mathcal{L}  \mathbf{h}^{(\ell)}_B \mathcal{P}_{\ell} \mathbf{h}^{(\ell)^*}_A = \sum_{\ell=1}^\mathcal{L} \bm{\mathcal{P}}_{BA}^{(\ell)},
\end{equation}
where $\mathbf{h}^{(\ell)}_A$ and $\mathbf{h}^{(\ell)}_B$ be the acoustic transfer function vectors from the $\ell^{\text{th}}$ source to group $\{A\}$ and $\{B\}$ microphones, respectively, and 
$\mathcal{P}_{\ell} \triangleq E\{|S_\ell|^2\}$.

Therefore, following the covariance subtraction, a set of independent $\hat{\mathcal{L}}$ sources can be used to practically generalize the ReTM, given as
\begin{equation}\label{eqn:retm_generalization}
\bm{\hat{\mathcal{R}}}_{AB} = 
\Big(\sum_{\ell=1}^{\hat{\mathcal{L}}} \bm{\mathcal{P}}_{AA}^{(\ell)} \Big)
\,  \Big( \sum_{\ell=1}^{\hat{\mathcal{L}}} \bm{\mathcal{P}}_{BA}^{(\ell)} \Big)^\dagger.
\end{equation}
Here, we note that a set of different active combinations of sources can be used to practically manipulate the ReTM estimation in \eqref{eqn:retm_generalization} via  
\eqref{eqn:phiAA_generalization} and \eqref{eqn:phiBA_generalization}.

The next section utilizes the proposed ReTM estimator to develop a more practically viable method for ReTM-based speaker separation.

%

\vspace{-0.4cm}
\section{Application into Speaker Separation}\label{sec:4}
\vspace{-0.3cm}

This section employs the proposed ReTM estimator to extract individual speakers from mixtures of multiple speech and noise sources, following the approach in \cite{wnmanamperiReTMSSS}.

Let us consider the number of concurrently active $\mathcal{L}_S$ speech and $\mathcal{L}_N$ background noise sources, where $\mathcal{L}=\mathcal{L}_S+\mathcal{L}_N$. In STFT domain, we denote $S_\ell^{(S)} $, $\ell = {1,\cdots,\mathcal{L}_S}$ and $S_n^{(N)}$, $n = {1,\cdots,\mathcal{L}_N}$ as the speech and background noise signals, respectively.

In the following, we extract the target speech of the $\ell^{th}$ speaker. Note that all speakers, $\ell = 1,\cdots,\mathcal{L}_S$, in the mixture can be similarly extracted. Let the target $\ell^{th}$ speech denotes as $S^{(S)}_\ell$ out of $L_S$ concurrent speakers. The rest of the undesired source signals including background noise can be grouped as 
$\mathbf{\bar{S}}_\ell = [S_1^{(S)} \ldots S_{\ell-1}^{(S)},S_{\ell+1}^{(S)}\ldots  S_{\mathcal{L}_S}^{(S)}, S_1^{(N)} \ldots S_{\mathcal{L}_N}^{(N)}]^T.$ Here we redefine the source signal vector as
$ \mathbf{S}(f,t) = [S^{(S)}_\ell \, \mathbf{\bar{S}}_\ell^T]^T$.

Let $\mathbf{\bar{H}}^{(\ell)}_A$ and $\mathbf{\bar{H}}^{(\ell)}_B$ be the acoustic transfer function matrices from all other sources except the $\ell^{\text{th}}$ speech source to group A and B microphones, respectively. Note that $\mathbf{H}_A = [\mathbf{h}^{(\ell)}_A ~ \mathbf{\bar{H}}^{(\ell)}_A]$ and $\mathbf{H}_B = [\mathbf{h}^{(\ell)}_B ~ \mathbf{\bar{H}}^{(\ell)}_B]$.

Denote $\bm{\bar{\mathcal{R}}}_{AB}^{({\ell})}$ be the ReTM of the combination of all sound sources except the $\ell^{th}$ target source. The separated speech signal of the $\ell^{th}$ speaker is then given by \cite{wnmanamperiReTMSSS} 
\begin{align}
\hat{\mathbf{S}}_\ell
& \triangleq  \mathbf{M}_A -  \bm{\bar{\mathcal{R}}}_{AB}^{({\ell})} \mathbf{M}_B, \label{eqn:desiredspeechl} \\
& =  [\mathbf{H}_A  - \bm{\bar{\mathcal{R}}}_{AB}^{({\ell})} \mathbf{H}_B ] [S^{(S)}_\ell \, \mathbf{\bar{S}}_\ell^T]^T, \label{eqn:est_speechvector} \\
 &= [\mathbf{h}^{(\ell)}_A  - \bm{\bar{\mathcal{R}}}_{AB}^{({\ell})} \mathbf{h}^{(\ell)}_B] S^{(S)}_\ell + \underbrace{[\mathbf{\bar{H}}^{(\ell)}_A  - 
\bm{\bar{\mathcal{R}}}_{AB}^{({\ell})} \mathbf{\bar{H}}^{(\ell)}_B}
_{\approx ~\mathbf{0}}
 ]
 \mathbf{\bar{S}}_\ell,  \nonumber \\
 &= [\underbrace{\mathbf{h}^{(\ell)}_A  - \bm{\bar{\mathcal{R}}}_{AB}^{({\ell})} \mathbf{h}^{(\ell)}_B}_{\text{distortion}}] S^{(S)}_\ell. \label{eqn:retm_sss}
\end{align}
We note that $\mathbf{\hat{S}}_\ell$ is a $ Q_A \times 1$ vector consists $Q_A$ copies of estimated target speech signal $S^{(S)}_\ell$. 

\setlength{\textfloatsep}{5pt} 
\setlength{\intextsep}{5pt}    
\setlength{\abovecaptionskip}{3pt} 
\setlength{\belowcaptionskip}{3pt} 

\begin{table*}[!h]
\caption{Speaker separation results for various SNR levels (SIR (dB)/SDR (dB)/STOI ($\%$)).} \label{table:simulations}
\vskip3pt
\centering
\begin{tabular}{|c|c|c|c|c|}
\hline
& \multicolumn{4}{|l|}{\centering \hspace{3cm} Method (SIR (dB)/SDR (dB)/STOI ($\%$))}\\ \cline{2-5}
SNR level & Unprocessed  &  ReTF - Oracle & ReTM (eq. \eqref{eqn:retm_phi}) & Proposed \\
\hline
$-20$ dB & $-4.15$/$-21.14$/$29.66$ & $27.29$/$\bm{4.84}$/$68.37$ & $\bm{30.37}$/$2.73$/$\bm{72.03}$ & $30.36$/$2.73$/$72.02$\\
\hline
$-15$ dB &  $-4.35$/$-17.69$/$31.21$ & $27.54$/$\bm{4.83}$/$68.48$ & $\bm{30.52}$/$2.71$/$71.80$ & $30.51$/$2.73$/$\bm{71.87}$\\
\hline
$-10$ dB & $-4.45$/$-13.59$/$33.95$ & $27.63$/$\bm{4.82}$/$68.51$ & $\bm{30.38}$/$2.52$/$\bm{71.78}$ & $30.35$/$2.50$/$71.77$\\
\hline
$-5$ dB & $-4.48$/$-9.74$/$37.86$ & $27.66$/$\bm{4.78}$/$68.49$ & $\bm{30.31}$/$2.67$/$71.85$ & $30.30$/$2.65$/$\bm{72.02}$\\
\hline
\hspace{0.35cm}$0$ dB & $-4.48$/$-7.44$/$41.44$ & $27.63$/$\bm{4.76}$/$68.45$ & $30.43$/$3.15$/$72.17$ & $\bm{30.46}$/$3.13$/$\bm{72.60}$\\
\hline
\end{tabular}
\end{table*}

In brief, \cite{wnmanamperiReTMSSS} proposed to extract the desired speaker from \eqref{eqn:desiredspeechl} with known segment boundaries of the undesired sources-only signals from the microphone recording to calculate the $\bm{\bar{\mathcal{R}}}_{AB}^{({\ell})}$ as 
\begin{equation}\label{eqn:retm_rhodesired}
     \bm{\bar{\mathcal{R}}}_{AB}^{({\ell})} \approx \bm{\bar{\mathcal{P}}}_{AA}^{({\ell})} \,  \Big( \bm{\bar{\mathcal{P}}}_{BA}^{({\ell})} \Big)^\dagger.
\end{equation}

Following the procedure in Section~\ref{sec:3}, i.e., adopting \eqref{eqn:phiS_AA}, \eqref{eqn:phiS_BA}, \eqref{eqn:retm_s}, and \eqref{eqn:retm_generalization}, we have \eqref{eqn:retm_rhodesired} as
\begin{align}
\bm{\bar{\mathcal{R}}}_{AB}^{({\ell})}
& = 
\Big
(
\sum_{\substack{\ell'=0 \\ \ell'\neq \ell}}^{\mathcal{L}_S} 
(\bm{\mathcal{P}}_{AA}^{(\mathcal{L}_N,\ell')} - \bm{\mathcal{P}}_{AA}^{(\mathcal{L}_N)}) +
\bm{\mathcal{P}}_{AA}^{(\mathcal{L}_N)}
\Big) \, \times
  \nonumber \\
& 
\Big( 
\sum_{\substack{\ell'=0 \\ \ell'\neq \ell}}^{\mathcal{L}_S}
(\bm{\mathcal{P}}_{BA}^{(\mathcal{L}_N,\ell')} - \bm{\mathcal{P}}_{BA}^{(\mathcal{L}_N)}) +
\bm{\mathcal{P}}_{BA}^{(\mathcal{L}_N)}
\Big)^\dagger. \label{eqn:retm_training}
\end{align}
where $\bm{\mathcal{P}}_{AA}^{(\mathcal{L}_N)}$ and $\bm{\mathcal{P}}_{AA}^{(\mathcal{L}_N,\ell')}$ are the covariance matrices of background noise-only signals, and background noise plus $\ell'$ speaker for $\ell' = 1,\ldots,\mathcal{L}_S$, respectively.

In teleconferencing, for a given meeting room with a fixed seating arrangement, both $\bm{\mathcal{P}}_{AA}^{(\mathcal{L}_N)}$ and $\bm{\mathcal{P}}_{AA}^{(\mathcal{L}_N,\ell')}$, $\ell' = 1,\ldots,\mathcal{L}_S$ can be recorded before the session begins. This algorithm does not require explicit pre-speech segments for each source, instead, it relies on the approximate independence of the speech sources, whereby their covariance contributions add up.


\vspace{-0.3cm}
\section{Experiments}\label{sec:5}
\vspace{-0.3cm}

This section presents experimental results using the proposed ReTM estimator for speaker separation with both simulated and real-life recordings at low ( $\leq 0$ dB) signal-to-background noise ratio (SNR) levels.


We evaluate speaker separation performance using (i) Signal-to-Interference Ratio (SIR), (ii) Signal-to-Distortion Ratio (SDR) (in BSS-Eval toolbox \cite{vincent2006performance}), and (iii) Short-Time Objective Intelligibility (STOI) \cite{taal2011algorithm}. The SNR is defined with respect to all sources in the mixture, whereas SIR is defined considering one speaker as the target signal and the rest of the interfering speakers as the noise signal.
To facilitate the interpretation of the results, we assume the availability of the oracle ReTF with its MVDR beamformer estimate \cite{wnmanamperiReTM2ReTF}. 

\vspace{-0.6cm}
\subsection{Simulated Environments}\label{sec:5.1}
\vspace{-0.3cm}


We utilize an open-source toolbox \cite{rirgen} to model the room impulse response (RIR) from sound sources to irregularly distributed microphones in a  $6 \times 7 \times 3$ m rectangular room ($T_{60} = 500$ ms). We consider $3$ speech sources, $2$ background noise sources, and $27$ microphones. 
We convolve the speech sources RIRs with both male and female speech utterances from the TIMIT dataset \cite{garofolo1993timit} and noise sources RIRs with wall air-conditioner noise, and vacuum noise. The received signals are ranged from $0$ to $-20$ dB SNR of background noise and added with $40$ dB SNR of white Gaussian noise at each microphone. Here, we define the SNR by averaging the SNR at each receiver over all $27$ receivers. The short ($60$ second)
recordings of background noise sources only, background noise sources plus each speaker are obtained for $\bm{\bar{\mathcal{R}}}_{AB}^{({\ell})}$ training from \eqref{eqn:retm_training}. These recordings are short-time-Fourier-transformed with an 8192-point window size that was long relative to the length of the RIR to satisfy the multiplicative transfer function \cite{avargel2007multiplicative}. We assign $Q_A = 10$, and $Q_B = 17$ number of receivers to group $\{A\}$ and $\{B\}$, respectively.
 
Table~\ref{table:simulations} depicts the speaker separation performance in various noisy environments. The results confirm that both ReTF- and ReTM-based estimators accurately separate all speakers in very low SNR levels. We observe the highest output SDR with the ReTF estimator, whereas the highest SIR and STOI values are achieved with the ReTM estimators, as the improved SIR leads to a slight reduction in SDR. However, we note that accurately estimating the ReTF is difficult in reverberant rooms with multiple noise sources. We also observe that the ReTM from \eqref{eqn:retm_phi} and the proposed ReTM estimator exhibit a similar performance.

\vspace{-0.6cm}
\subsection{Real-life Environments}\label{sec:5.3}
\vspace{-0.5cm}

\setlength{\textfloatsep}{5pt} 
\setlength{\intextsep}{5pt}    
\setlength{\abovecaptionskip}{3pt} 
\setlength{\belowcaptionskip}{3pt} 

\begin{table}[!h]
\centering
\caption{
Speaker separation results at low SNR $= \{-10, 0\}$ dB for real recordings.
The values are increments with respect to the unprocessed signals and averaged across speakers ($\Delta$SIR (dB)/$\Delta$SDR (dB)/$\Delta$STOI ($\%$))
}
\label{tab:table2}
\begin{tabular}{|c|c|c|}
\hline
SNR & \multicolumn{2}{|l|}{\centering Method ($\Delta$SIR (dB)/$\Delta$SDR (dB)/$\Delta$STOI ($\%$)) }\\ \cline{2-3}
level  &  ReTF - Oracle & Proposed \\
\hline
-10 dB  & ${4.70}$/${0.15}$/$1.78$ & $\bm{11.42}$/$\bm{8.13}$/$\bm{19.93}$\\
\hline
0 dB & $20.15$/$1.4$/$3.46$ &  $\bm{26.12}$/$\bm{4.9}$/$\bm{31.78}$\\
\hline
\end{tabular}
\end{table}


Next, we examine the performance of the proposed method in real-life scenarios. The experimental recordings are measured in an office room at the Australian National University with dimensions  $2.95 \times 6 \times 3.03$ m, and $T_{60} \approx 630$ ms. We consider $5$ loudspeakers, including $3$ speakers, $2$ background noise sources (fan and room air cooler), and $15$ randomly distributed microphones over the room. We assigned $Q_A = 5$, and $Q_B = 10$ number of receivers to group $\{A\}$ and $\{B\}$, respectively.
We examined both the `Proposed' and `ReTF - Oracle' estimators to separation performance in Table~\ref{tab:table2}. Again, the proposed method consistently improves the average SIR, SDR, and STOI over the mixture signals for all three speakers compared to the baseline. The `Proposed' estimator performs as anticipated by simulation analysis conclusions, with slightly lower results due to unavoidable practical errors.

%

\vspace{-0.5 cm}
\section{Conclusion}\label{sec:5}
\vspace{-0.3cm}

This paper proposed a novel
method to estimate the ReTM using covariance matrices in a noisy, reverberant environment with multiple speech and noise sources. The method uses covariance subtraction to estimate the ReTM for different subgroups of sound sources. We showed that the ReTM is not necessarily additive. We generalized the ReTM estimation for a select set of independent sources to manipulate the covariance matrices between microphone groups along with covariance subtraction. We evaluated this method on a speaker separation application using both simulation and real-life recordings. Extensive experimental study confirmed that this method achieves strong separation results under very low SNR conditions. 
In the future, we plan to utilize this method to sound zone control problem.


\ninept
\bibliographystyle{IEEEbib}
\bibliography{paper}

\end{document}